\documentclass[doublecol]{epl2} 

\title{No steady state flows below the yield stress. A true yield stress at last?}
\shorttitle{No steady state flows below the yield stress.} 

\author{Peder~C.~F. M{\o}ller\inst{1} \and Abdoulaye Fall\inst{2} \and Daniel Bonn\inst{1,2}}

\shortauthor{P. M{\o}ller \etal}
\institute{                    
  \inst{1} \'{E}cole Normale Sup\'{e}rieure, Laboratoire de
Physique Statistique, Paris, F-75231 France
  \inst{2} van der Waals-Zeeman Institute, University of
Amsterdam, 1018 XE Amsterdam, The Netherlands
}
\pacs{83}{Rheology}

\date{\today}

\abstract{
For more than 20 years it has been debated if yield
stress fluids are solid below the yield stress or actually flow; whether true yield stress fluids exist or not. Advocates of
the true yield stress picture have demonstrated that the effective
viscosity increases very rapidly as the stress is decreased
towards the yield stress. Opponents have shown that this viscosity
increase levels off, and that the material behaves as a
Newtonian fluid of very high viscosity below the yield stress. In
this paper, we demonstrate experimentally (on four different
materials, using three different rheometers, five different geometries, and two different measurement methods) that the low-stress
Newtonian viscosity is an artifact that arises in non steady state experiments. For measurements as long
as $10^4$ seconds we find that the value of the 'Newtonian
viscosity' increases indefinitely. This
proves that the yield stress exists and marks a sharp transition
between flowing states and states where the steady state viscosity
is infinite -a solid!
}

\begin{document}

\maketitle

\section{Introduction}

Yield stress materials can be either 'fluid' or 'solid'; a typical
example is toothpaste, that flows (i.e., is fluid) when pushed out
of the tube but no longer flows (is solid) under the influence of
gravity while posed on a toothbrush. Such materials are of
paramount importance for a large number of applications; examples
are concrete, oil drilling fluids, granular and building
materials, cosmetic products, and foodstuffs. For all these yield
stress fluids, it is important to understand their flow
characteristics and notably the yield stress to correctly predict
their flow behavior.

The canonical yield stress picture assumes that the material is solid
until a critical shear stress is exceeded. Above that stress the material
yields and subsequently flows. The Bingham or the Herschel-Bulkley (H-B) models are widely used to
such behavior; the
latter reads: $\sigma=\sigma_y+\alpha {\dot \gamma}^n$, where
$\sigma$ is the shear stress (the subscript $y$ indicates the
yield stress) and ${\dot \gamma}$ the velocity gradient or shear
rate; $a$ and $n$ are adjustable model parameters. This model can
successfully describe the flow behavior of simple yield stress
fluids such as carbopol, emulsions, foams, etc.\ over large ranges
of shear rates. \emph{Thixotropic} yield stress fluids, where the
flow changes the material structure which in turn leads to a varying viscosity and yield stress,
have been reviewed recently in \cite{Moller:2006p768} and are not considered here.

\begin{figure}[ht]
\centering
\begin{minipage}[c]{0.16\textwidth}
\onefigure[width=\textwidth]{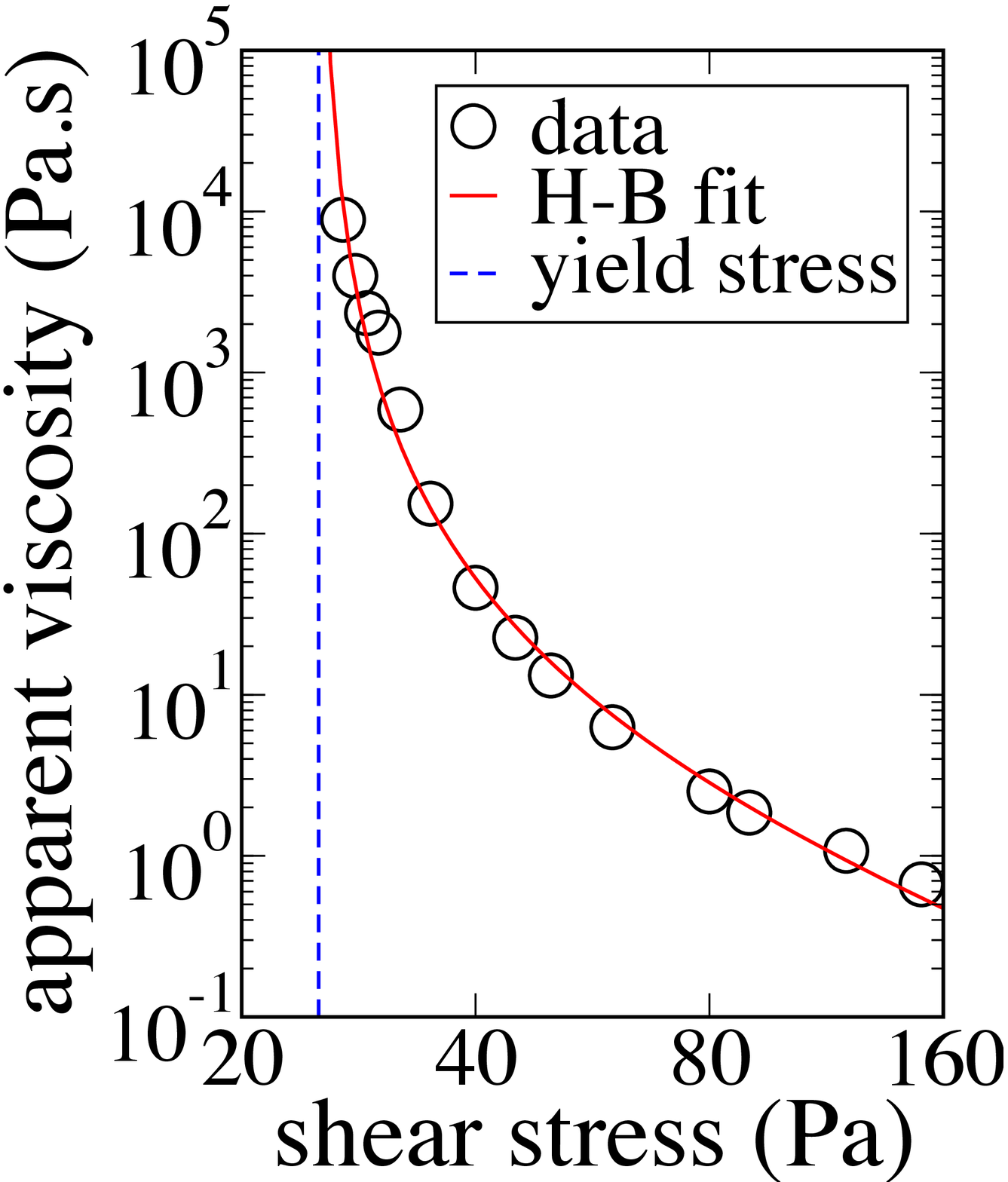}
\end{minipage}
\begin{minipage}[c]{0.31\textwidth}
\onefigure[width=\textwidth]{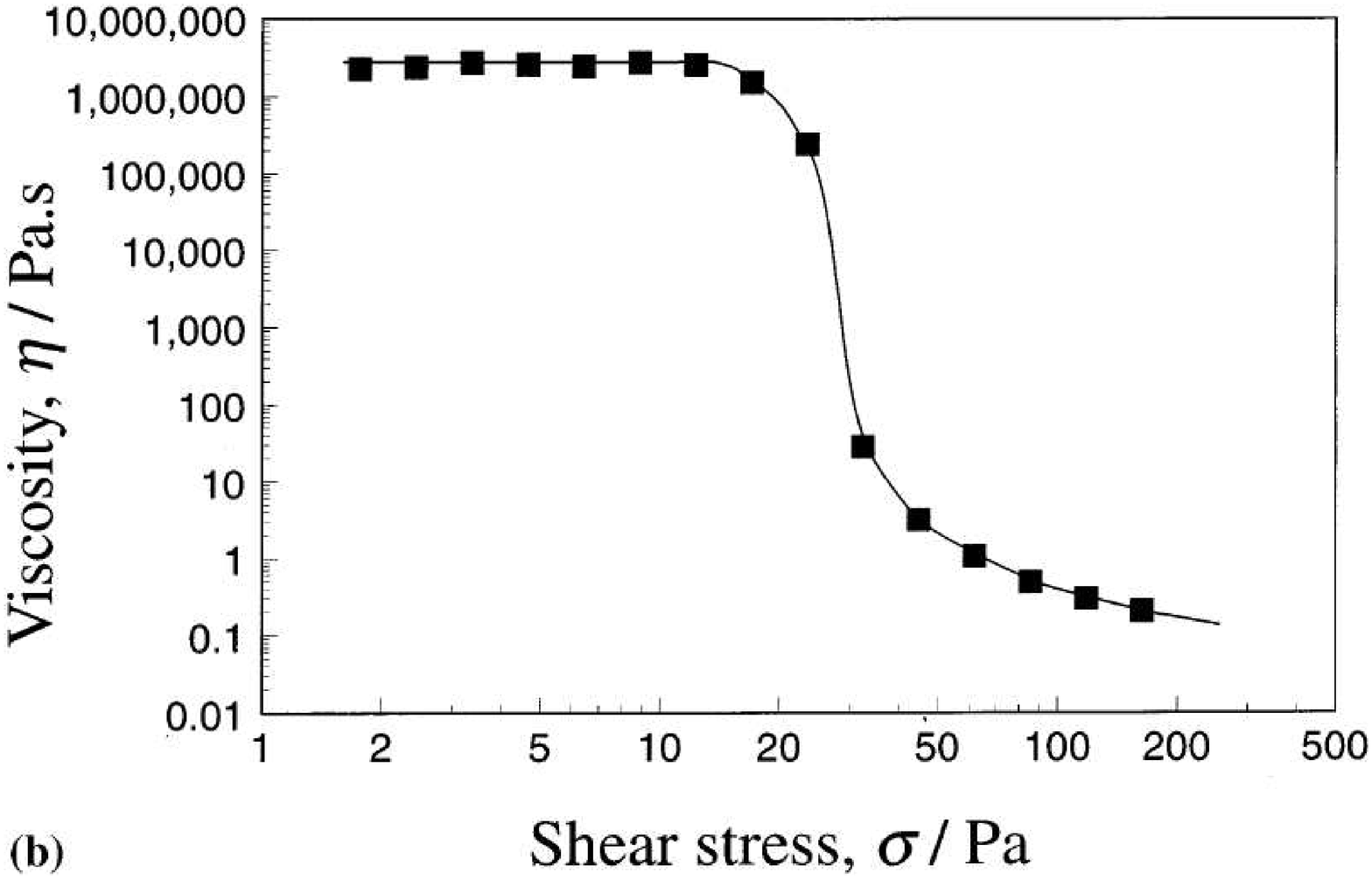}
\end{minipage}
\caption{
\strut\vspace{-1.8cm} \newline\strut  \hspace{1.0cm} \textbf{A}\hspace{3.3cm}\textbf{B}
\vspace{1.05cm}\newline\strut \hspace{1.1cm}
\textbf{A}
The Herschel-Bulkley model provides a good fit to the viscosity of a carbopol sample (0.2~\%${_{mass}}$ at $pH=7$) as $\sigma$ is lowered towards $\sigma_y$.
\textbf{B}
Measurements on
an identical Carbopol sample
apparently demonstrate the
existence of a Newtonian limit below the 'yield stress' \cite{Barnes:1999p8}. This figure is
representative of the figures claimed to demonstrate that yield stress
materials flow below the yield stress. With kind permission from Elsevier.
\label{barnes-carbopol}
}
\end{figure}

For our discussion, the key prediction of the H-B and
other generic yield stress models is that the viscosity (defined
as the ratio of the stress and shear rate) diverges continuously
when the yield stress is approached from above. Hence, below the
yield stress, the viscosity is infinite, and the material behaves
as a solid. Fig.~\ref{barnes-carbopol}\textbf{A}, depicts a
H-B fit to the viscosity of a carbopol sample, from which it appears that the viscosity
indeed diverges as $\sigma$ is decreased towards $\sigma_y$.

However, for a number of years there has been a controversy as to
whether or not the yield stress marks a transition between such a
'solid' and a 'liquid' state, or if it instead marks a transition between two liquid
states with very different viscosities. That is, if true
yield stress fluids actually exist. Probably the earliest work
that seriously questions the solidity of yield stress fluids below
$\sigma_y$ is a 1985 paper by Barnes and Walters
\cite{Barnes:1985p323}. They show data on carbopol samples
apparently demonstrating the existence of a finite viscosity
 plateau (i.e.\ Newtonian behavior) at very low shear stresses - rather than an
infinite viscosity below the yield stress. In his review paper on
the subject "The yield stress - a review or
'$\pi\alpha\nu\tau\alpha\ \rho\epsilon\iota$' - everything
flows?", Barnes presents numerous flow curves with viscosity
plateaus at low stresses \cite{Barnes:1999p8}. One of these curves
is reproduced in Fig.~\ref{barnes-carbopol}\textbf{B} where the
viscosity plateau appears very convincing.

Following these publications, a series of papers appeared
discussing the proper definition of yield stress fluids, whether
they existed or not, and how to demonstrate either way,
e.g.~\cite{Evans:1992p11,Spaans:1995p7,Hartnett:1989p1,Schurz:1990p12,Barnes:2007p219}.
The outcome of this debate has been that the rheology and soft matter communities
presently hold two coexisting yet conflicting views: (1) the
yield stress marks a transition between a liquid state and a solid
state, and (2) the yield stress marks a transition between two
fluid states that are not fundamentally different - but with very
different viscosities.

In this Letter, we reproduce the experiments used to demonstrate
Newtonian limits below the yield stress and we too find the
apparent viscosity plateaus at low stresses. However, we show that such
curves are artifacts that arise when falsely concluding that a
steady state has been reached. We find that, for imposed
stresses below the yield stress and for measurement times at least
as long as $10^4$ seconds, apparent viscosities increase with time
and show no signs of approaching a constant value. This reveals that below
the yield stress, the material is not fluid but solid.

In order to experimentally examine the nature of the yield stress
transition as well as to test the generality of our findings, we used
four different yield stress fluids, three different rheometers,
five different measurement geometries and two different
measurement protocols. The materials used were; carbopol, a
commercial hair gel (Gatsby Water Gloss/Wet Look Soft), a
commercial shaving foam (Gillette Foamy Regular), and a cosmetic
water-in-oil emulsion. Carbopol is being used on a huge industrial
scale as a thickening/yield stress agent but it is also a favorite
'model' simple yield stress fluid for researchers
\cite{Barnes:1985p323,Barnes:1999p8,Piau:2007p1,Uhlherr:2005p101,Atapattu:1995p245,Atapattu:1990p31}.
Carbopol powder consists of small 'sponges' of cross-linked
polyacrylic acid resins that, under neutral $pH$ swell so
enormously in water that concentrations of carbopol even below
$0.1~\%_{{mass}}$ are sufficient for the particles to jam
together to form a yield stress fluid \cite{Piau:2007p1}. The
carbopol sample used for the experiments presented here was
prepared by mixing 0.2~\% by mass of carbopol Ultrez U10 grade and
ultra-pure water thoroughly for at least an hour which forms an
opaque, viscous fluid with no yield stress. The $pH$ is then
adjusted to 7 by adding NaOH, leading to a clear material with a yield stress. After pH
calibration and thorough mixing, the fluid was left for at least a
day and mixed again, before any measurements were done. The
commercial hair gel containing carbopol, water and triethanolamine
as a stabilizing agent was also mixed prior to use to assure
homogeneity. The foam can was shaken well before dispensing the
foam, and since the very first and the very last foam from the can
differs from the rest, only foam from the bulk was used for the
experiments. The water-in-oil emulsion (IB80, Fabre, France) was
described previously in \cite{Bertola:2003} and remains stable
under shear. The size of the water droplets measured by microscopy
was around $7~\mu$m. For all systems, the characteristic time
scales of possible changes in the material properties (e.g.,
coarsening, for the foam and emulsion) were much larger than the
duration of our tests.

\begin{figure}[ht]
\centering
\begin{minipage}[c]{0.2385\textwidth}
\centering
\onefigure[width=\textwidth]{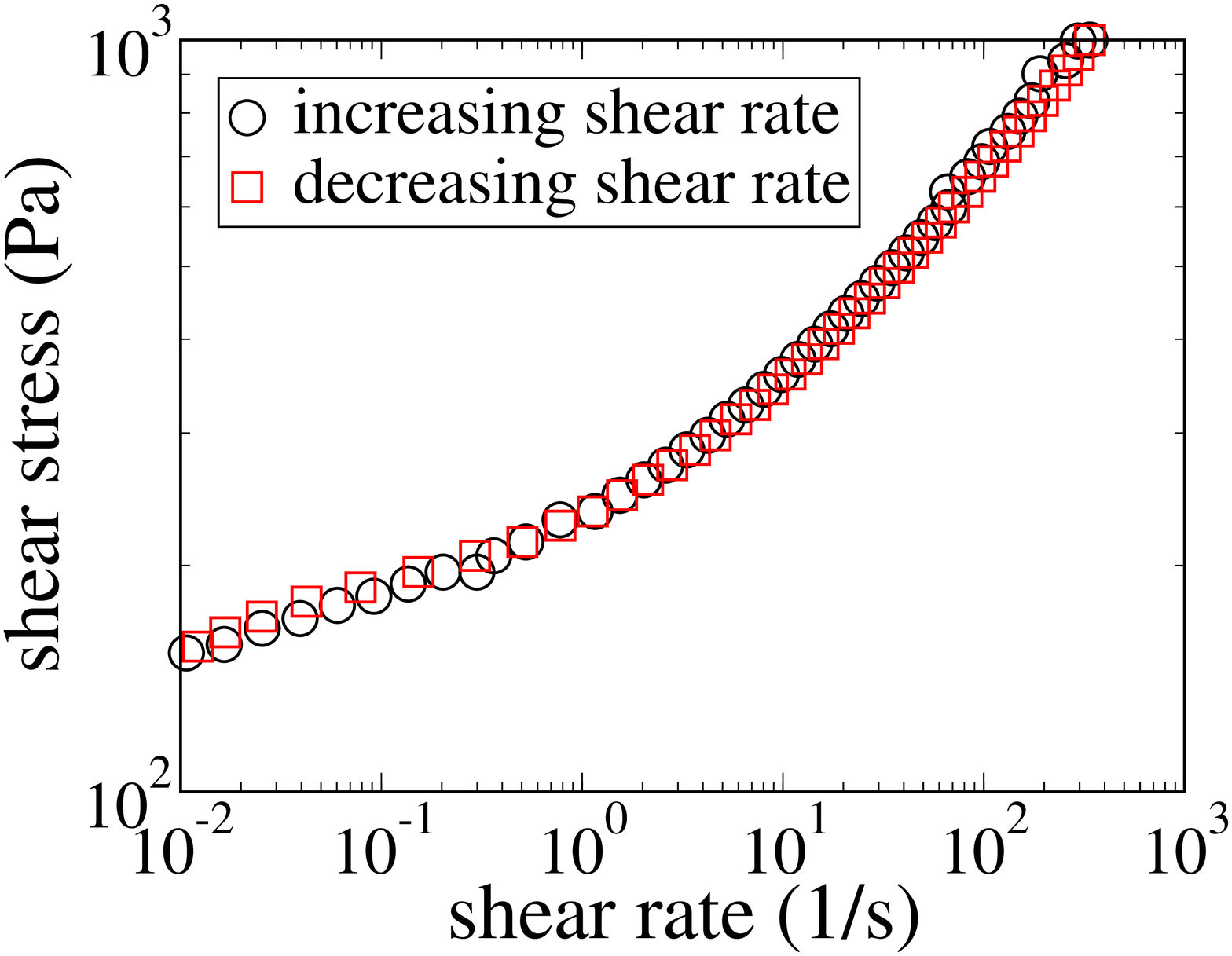}
\end{minipage}
\centering
\begin{minipage}[c]{0.2385\textwidth}
\centering
\onefigure[width=\textwidth]{carbopol-updown.eps}
\end{minipage}
\caption{ \strut\vspace{-1.7cm} \newline\strut  \hspace{3.5cm}
\textbf{A}\hspace{4.2cm}\textbf{B} \vspace{0.95cm}\newline\strut
\hspace{1.1cm} A material is normally tested for thixotropy by
imposing an up-down rate-sweep, where the shear stress is measured
while the shear rate is progressively increased and then decreased. If
the material is thixotropic, the viscosity is time-dependent
leading to a hysteresis effect: the increasing and decreasing
curves do not overlap. Both the emulsion (\textbf{A}), and the
carbopol (\textbf{B}) are seen to be entirely non-thixotropic, as
is the also case for the foam and the hair gel (data not shown).
Carbopol data from measurements with both the 25 mm radius cone-plate geometry
and the 14.4 mm radius vane-cup geometry collapse nicely, 
demonstrating that consistent results are obtained even with completely different geometries.
Since wall slip would have caused two geometries with very different gap sizes to yield different results at low shear rates, 
this collapse for data from geometries with 0.2 mm and 3.4 mm gaps respectively demonstrates that we avoid wall slip.
\label{nonthix}}
\end{figure}

The measurements on the carbopol and the hair gel presented here
were done using a vane-cup geometry (9~mm inner radius, 13.25~mm
outer radius and 52~mm height) in a controlled stress rheometer
(Stresstech from Rheologica). The measurements on the foam were
made with another vane-cup geometry (11~mm and 14.4~mm radii, and
16~mm height) on a Paar Physica MCR 300 rheometer, while the
measurements on the emulsion were done with a cone-plate geometry
with a 25~mm radius and a $1^{\circ}$ opening angle on the MCR
300. These measurements were double-checked on a Haake RS150
instrument. Besides the measurements shown here we also performed
measurements with 12.5~mm radius cone-plate geometries with
opening angles of $0.25^{\circ}$ and $1^\circ$. All geometries
had roughened walls in order to prevent wall slip effects
during measurements. All results presented here are independent of
the rheometer and geometry, indicating that wall slip was
successfully prevented \cite{Bertola:2003} (see Fig.~\ref{nonthix}B).
Shear banding in non-thixotropic yield stress fluids (as opposed to thixotropic yield stress fluids)
 happens only if there is
significant stress variation within the geometry. So the fact that our measurements are independent of whether a vane-cup (with a large stress variation) or a cone-plate 
(with a neglegctable stress variation) is used demonstrate that shear banding does not affect our conclusions \cite{Moller:2008p041507}.

\begin{figure}[ht]
\centering
\begin{minipage}[c]{0.2385\textwidth}
\centering
\onefigure[width=\textwidth]{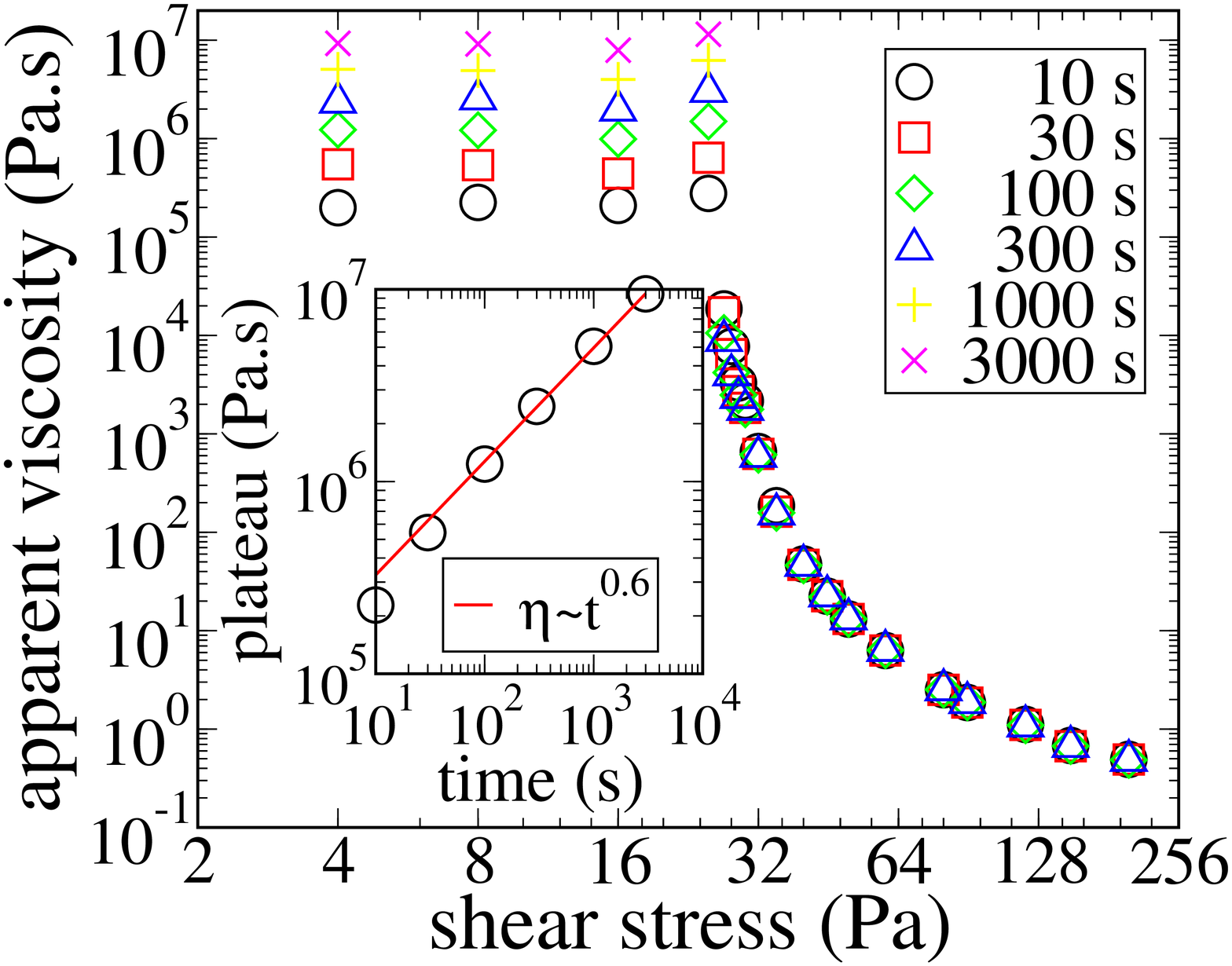}
\end{minipage}
\centering
\begin{minipage}[c]{0.2385\textwidth}
\centering
\onefigure[width=\textwidth]{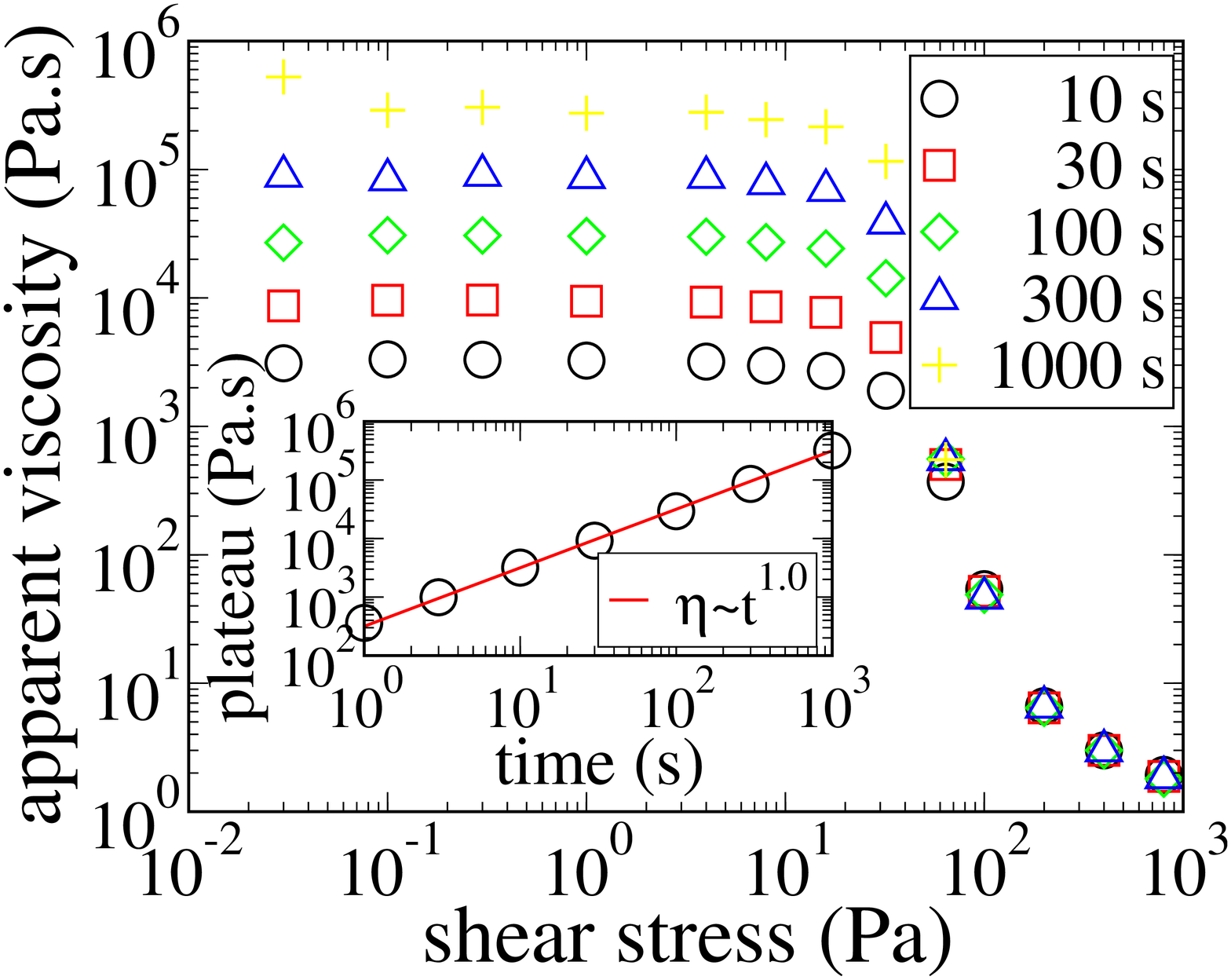}
\end{minipage}
\centering
\begin{minipage}[c]{0.2385\textwidth}
\centering
\onefigure[width=\textwidth]{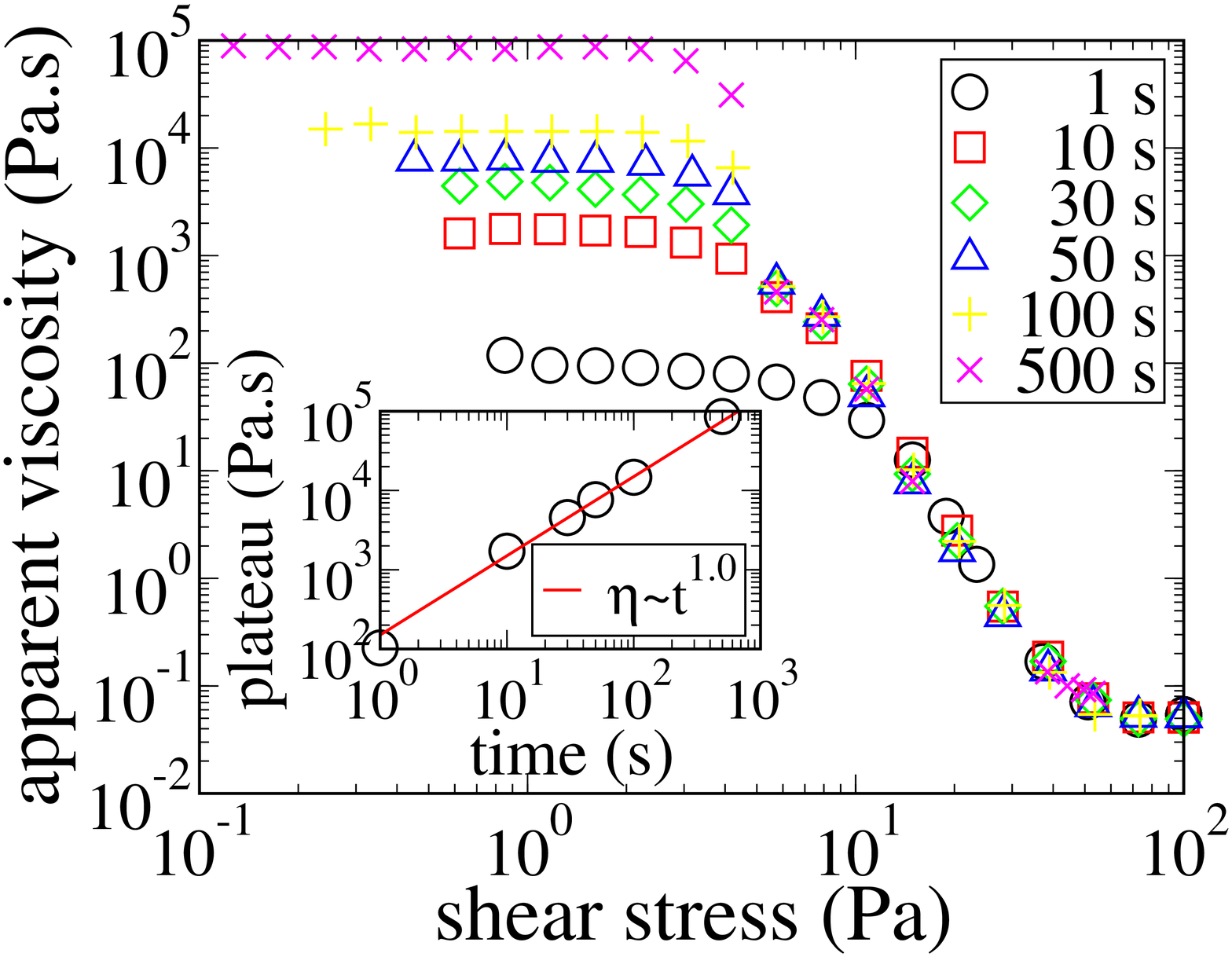}
\end{minipage}
\centering
\begin{minipage}[c]{0.2385\textwidth}
\centering
\onefigure[width=\textwidth]{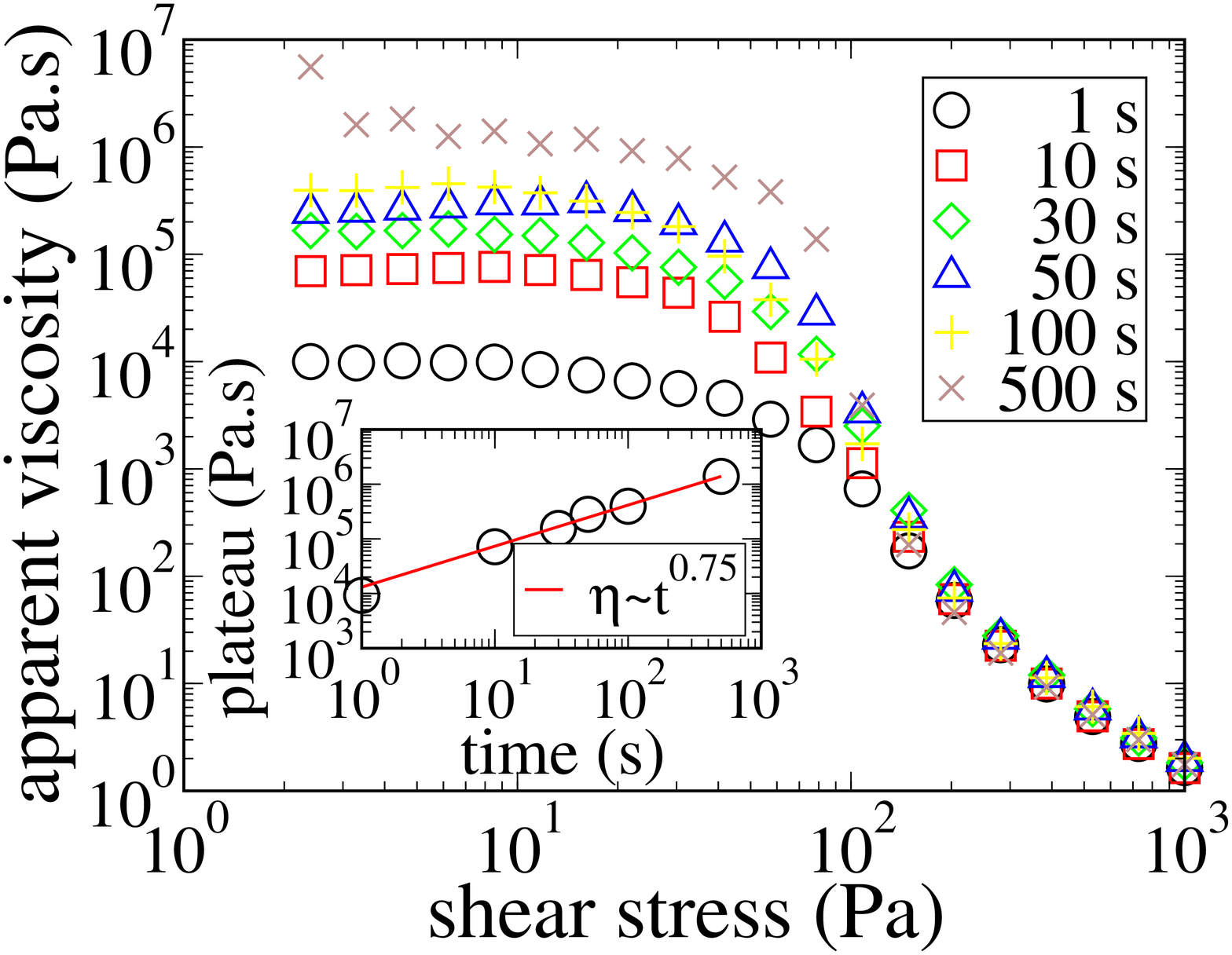}
\end{minipage}
\caption{ \strut\vspace{-5.5cm} \newline\strut  \hspace{3.5cm}
\textbf{A}\hspace{4.2cm}\textbf{B} \vspace{3.0cm}\newline\strut
\hspace{3.65cm}\textbf{C}\hspace{4.2cm}\textbf{D}
\vspace{1.4cm}\newline\strut \hspace{1.1cm}
Reproductions of
measurements like the one on Carbopol in
Fig.~\ref{barnes-carbopol}\textbf{A}. For each material and each
measurement time the resulting curve resembles that in
Fig.~\ref{barnes-carbopol}\textbf{A}, but the values of the
plateaus increase with measurement time. The insets show that the
plateaus increase as power laws with time with exponents in the
range 0.6-1.0. \textbf{A} The 0.2~\% carbopol sample. \textbf{B}
The hair gel. \textbf{C} The foam. \textbf{D} The emulsion.
\label{plateaus}}
\end{figure}

The main issue is whether the viscosity is finite or infinite
below the yield stress. This means resolving very small shear rates
which classical rheology do better than velocimetry. 
We therefore made classical creep tests: the shear
stress is imposed and the resulting global shear rate is recorded. 
We use the term \emph{apparent} viscosity for the stress divided by 
the instantaneous shear rate to underline that this quantity might be 
time dependent and hence not a 'true' viscosity - even for a given stress value. The resulting 
apparent viscosity \emph{vs.}~shear stress curves are shown in Fig.~\ref{plateaus}. For
all samples, low-stress Newtonian viscosity plateaus very similar
to the ones reported in the literature are found. However, while
the measurements for all measurement times collapse at stresses
higher than the ``yield stress'', the apparent viscosity values
are time-dependent below it: the viscosity value of the Newtonian
viscosity plateau increases with the measurement time. It is clear
that, when seen individually, the curves for the four different
materials shown in Fig.~\ref{plateaus} can easily be misleading 
since such nice plateaus are normally seen only for Newtonian,
time independent materials. In reality, the viscosity value
of the 'Newtonian plateau' increases without bound. Although the
trend is identical, the four different systems give slightly
different exponents for the time evolution of the viscosity: $\eta
\sim t^\mu$, with $\mu\in [0.6,1]$, as shown in the insets in
Fig.~\ref{plateaus}.

To demonstrate the robustness of the phenomenon two different test
methods were used. \emph{Method 1} was used on the carbopol and
the hair gel sample. Here, a constant stress is imposed and the
apparent viscosity is recorded as function of time for each
stress. \emph{Method 2} was used on the foam and emulsion. Here,
the imposed stress was increased stepwise, with the duration of
each step (over which the apparent viscosity is averaged) varying
from series to series. As is evident from Fig.~\ref{plateaus},
independently of the measurement method, material, rheometer, and measurement geometry, we retrieve the apparent
plateaus at low stresses, that evolve with time and thus do not
correspond to steady state viscosity values.

We believe that the viscosity plateaus in the litterature are in 
general caused by the phenomenon we demonstrate here, and 
that they are therefore not to be trusted unless data demonstrating that the  
measurements represent a steady state are provided. We have demonstrated that 
measurement times much longer than 10,000 seconds are needed to reach a steady state
for typical yield stress fluids, and we belive that authors would surely have 
commented on the need for so long measurement times had they been aware of it. 
For the specific case of carbopol, the data in Fig.~\ref{barnes-carbopol} 
agree perfectly with our measurements of carbopol after five minutes of stress 
in Fig.~\ref{plateaus}\textbf{A}. Five minutes is normally a reasobale time to 
reach steady state in a rheology experiment, and in combination with the nice newtonian plateaus, 
this is perhaps the reason why previous authors have not questioned whether their data corresponded to a true steady state.

\label{YSResults}

\begin{figure}[ht]
\centering
\begin{minipage}[c]{0.28\textwidth}
\centering
\onefigure[width=\textwidth]{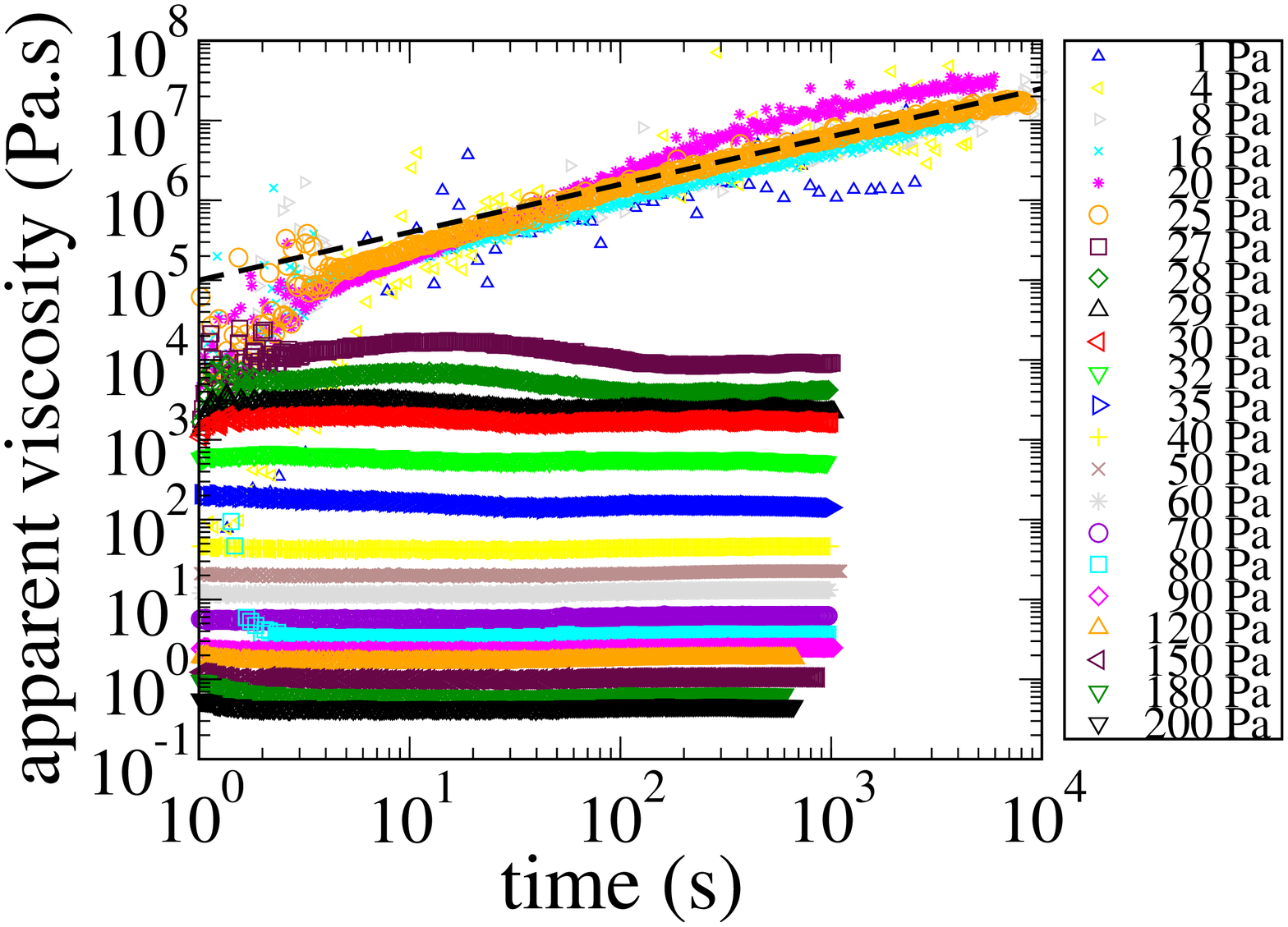}
\end{minipage}
\centering
\begin{minipage}[c]{0.196\textwidth}
\centering
\onefigure[width=\textwidth]{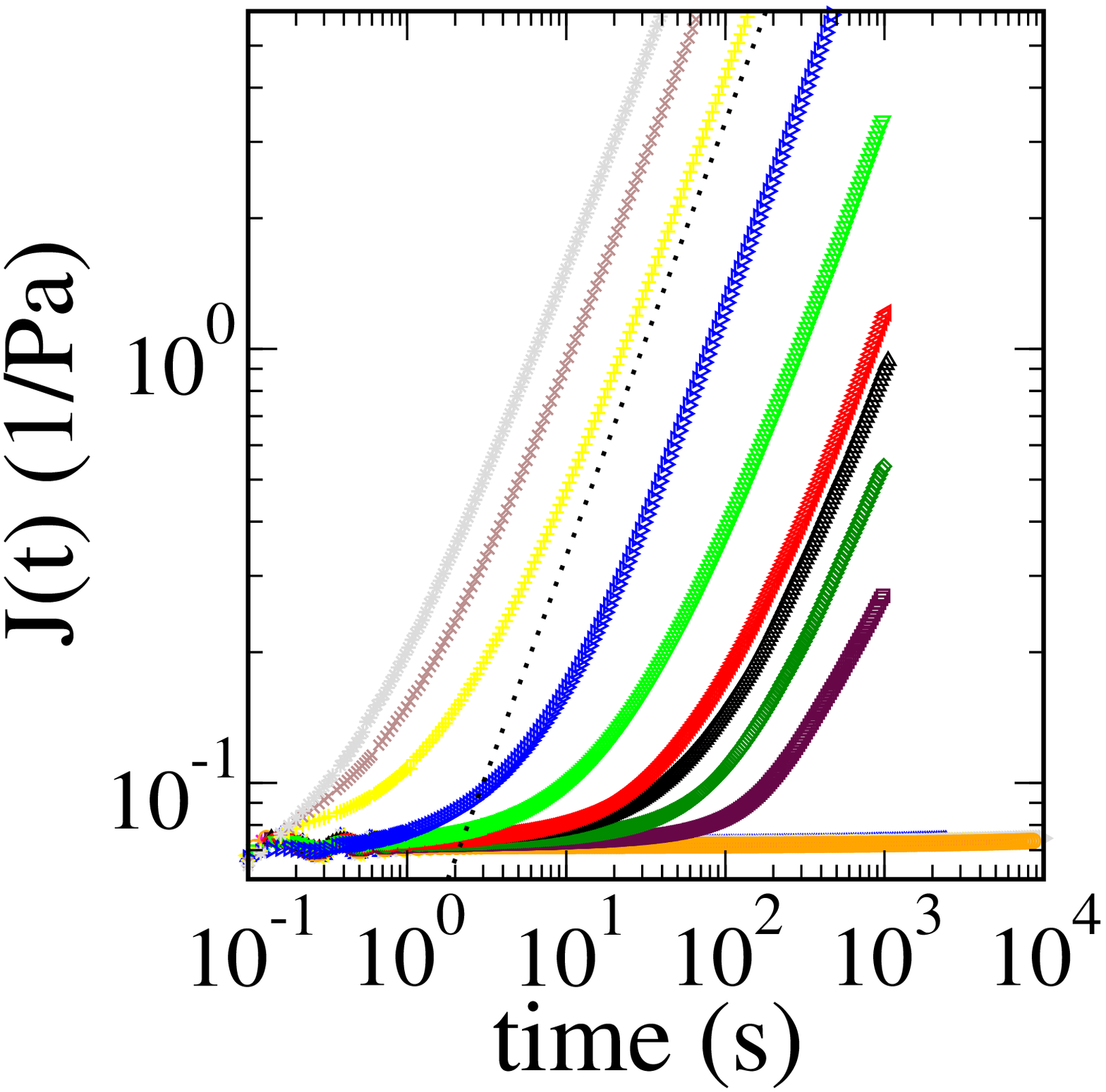}
\end{minipage}
\caption{\strut\vspace{-2.0cm} \newline\strut  \hspace{3.5cm}
\textbf{A}\hspace{4.2cm}\textbf{B} \vspace{1.2cm}\newline\strut
\hspace{1.1cm} 
Time evolution of 0.2~\% carbopol at $pH=7$.
\textbf{A} The apparent viscosity is time-independent but stress-dependent at $27~Pa$ and above, but stress-\emph{in}dependent and increasing with time 
(close to $\eta\sim t^{0.6}$ as indicated by the dashed line) at $25~Pa$ and below. 
 This behavior is what gives rise to the stress plateaus. \textbf{B} A creep compliance curve of the same data, showing the time evolution of $J(t)$ - the total deformation divided by the imposed stress. 
 At large $t$ a fluid will have a slope of 1 (indicated by the dotted line), while a solid will have a slope of 0. At stresses of 27 Pa and above, slopes are nearly 1, while they are nearly 0 at 25 Pa and below.
\label{visc-time}}
\end{figure}

In Fig.~\ref{visc-time} the qualitative difference between
imposing a stress below and above the yield stress of the 0.2~\%
carbopol sample is clearly seen. For stresses at or above $27~Pa$
the viscosity quickly reaches a steady state, but stresses at or
below $25~Pa$ (corresponding to the last point in the viscosity
plateau of 0.2~\% carbopol) the viscosity keeps increasing in time as $\eta\sim t^{0.6}$
for times even longer than $10^4$ seconds. Since evidently no
steady state is observed below the yield stress one should not
take the instantaneous, apparent viscosity at \emph{any} point in
time to be proof of a high-viscosity Newtonian limit at low
stresses for these materials.

It should be underlined that measurements for each material
presented here were made on the very same sample. Two different
samples from the same batch gave identical results. Near the end
of a series of measurements, initial measurement points were
reproduced to make sure that evaporation, aging or material
degradation had not changed the properties of the material. After
loading the sample into the geometry and between constant, imposed
stress measurements, the material is allowed to relax at zero
applied stress for half an hour to partially relieve internal
stresses built up during loading or during the previous
measurement. Thus, the systems do not spontaneously 'age' in the
sense that at rest the mechanical properties do not evolve in
time.

To the contrary, when subjected to a stress below the yield
stress, the viscosity increase in time does suggest aging. Other
soft glassy materials however, typically show aging at rest (an
increase of the viscosity with time) and 'rejuvenation' under
stress, i.e., a viscosity decrease when the system is made to
flow. \cite{Bonn:2002p312, Bonn:2002p363,  Sollich:1998p510,
Sollich:1997p331, Lequeux:2001p461, Berthier:2000p676,
Varnik:2004p2479, Hebraud:1998p2483, Evans:1999p2484}. For the
carbopol, hair gel, foam, and emulsion, an aging under stress but not at rest
is observed.

The spontaneous aging at rest observed for many soft glassy
materials is mostly a thermally driven process, for instance in a colloidal gel where Brownian
forces move particles around until
they aggregate into a microstructure that resists flow. For
glassy systems the thermal noise moves the system through a
succession of continuously deeper local minima of the free energy
landscape which become continuously harder to escape from,
implying a higher relaxation time / viscosity. Rejuvenation by the
flow can then be looked upon as the shear in the system playing
the role of a very high effective temperature, causing the
microstructures to break up, or the glassy system to escape from
the a local free energy minimum. Such a competition between aging
and shear rejuvenation is called thixotropy in rheological terms
\cite{Moller:2006p768}. However, carbopol, foams and emulsions are
normally considered to be \emph{non}-thixotropic systems.
Explicit measurements of the thixotropy are shown in
Fig.~\ref{nonthix} and show that this is indeed the case. In
general, the thixotropy is quantified using the thixotropy index,
which is the difference in area under the flow curves of
increasing and decreasing stress; as is evident from
Fig.~\ref{nonthix}, this is zero to within the experimental
accuracy for the systems used here.

The question is therefore how to understand this 'aging under stress'
of a non-thixotropic system? The absence of aging at rest is easy
to understand. For the systems used here, the thermal agitation is
not nearly enough to reorganize the arrangement of droplets,
bubbles, or carbopol 'sponges', since due to the large size of the
particles or bubbles, the free energy barriers for structural
reorganization are much larger than $kT$. This is why in the
experiments, there is no detectable aging of these systems when
they are at rest. However, if a stress is imposed on the material
the energy landscape is tilted, facilitating reorganizations in
the direction of the imposed stress. If the imposed stress is so
large that the system can escape from all of its local minima, the
material will keep flowing at this stress and a 'simple' viscous
flow results. However, if the imposed stress is large enough to
cause some structural reorganization but insufficient for the
system to explore all of the available phase space, the system
will age just as described above for the soft glassy systems, the
imposed stress playing the role of an increased effective
temperature. This is very similar to a loose granular material 
that does not compact (age) when left at rest, but does when subject
to small shear rates or vibrations. That small stresses can actually 
facilitate aging processes rather than rejuvenation processes has 
previously been observed for some thixotropic systems (that age even 
without any stress) and has been termed 'overaging' 
\cite{Viasnoff:2002,Lacks:2004,Warren:2008}. It has until now not been 
 observed in non-thixotropic fluids.

In conclusion we have shown that the many viscosity \emph{vs.}\
stress curves in the literature showing Newtonian limits at low
stresses should not be trusted unless data proving that the points
represent a steady state are provided. This is because, below the
yield stress, the viscosity is independent of stress but increases
as a power law in time after the stress has been imposed, so that
recording the viscosity after say 100 seconds will give one
viscosity plateau while data at 1,000 seconds will give another.
So the instantaneously measured viscosity does not represent a
steady state and should not be reported as a proof for a Newtonian
limit for the material at low stresses. We have demonstrated this
increase of the viscosity with time for four different typical
simple yield stress systems and it persists for at least as long
as 10,000 seconds. Our findings are well explained qualitatively by assuming that aging of the
system happens at experimental time scales only if a small but
non-zero stress is applied to the system: at zero stress the
thermal agitations are too weak to cause reorganizations, and at
too high stress the materials constantly flows and the system is
rejuvenated. The yield stress is the limit between 'aging
stresses' and 'rejuvenation stresses': above it the material flows
with a time-independent viscosity, and below it the system ages
and the steady state viscosity is infinite. Thus, the yield stress
is real, and it marks a transition between a flowing and a 'solid'
state.
\acknowledgments
We thank M.~A.~J. Michels and P. Coussot for very helpful
discussions.

\end{document}